\documentclass[smallextended]{svjour3}       
\smartqed  
\usepackage{graphicx}
\newcommand{\beq}{\begin{equation}}
\newcommand{\eeq}{\end{equation}}
\newcommand{\beqn}{\begin{eqnarray}}
\newcommand{\eeqn}{\end{eqnarray}}
\newcommand{\bmath}{\begin{subequations}}
\newcommand{\emath}{\end{subequations}}

\usepackage{hyperref}
\hypersetup{
    colorlinks=true,
    linkcolor=blue,
    filecolor=blue,      
    urlcolor=blue,
}

\begin{document}

\title{Superconducting materials: the w{\it hole} story \\
{\it Dedicated to Ted Geballe on his 100th birthday}
}


\author{J. E. Hirsch 
}


\institute{J. E. Hirsch \at
              Department of Physics, University of California, San Diego,
La Jolla, CA 92093-0319\\
              Tel.: +18585343931\\
               \email{jhirsch@ucsd.edu}           
}

\date{Received: date / Accepted: date}

\maketitle

\begin{abstract}
Ted Geballe has contributed  enormously to the knowledge of superconducting materials during an illustrious
scientific career spanning seven decades, encompassing groundbreaking discoveries and  studies of both so-called conventional and unconventional superconductors.
On the year of his 100th birthday  I would like to argue that all superconducting materials that Ted investigated, as well as those he did not, have
one thing in common that is not generally recognized: hole carriers. This includes
$PbTe$ doped with $Tl$, for which Ted has proposed that superconductivity is driven by negative-U pairing. I will discuss why  hole
carriers are necessary for a material to be a superconductor, and the implications of this for the understanding of the fundamental physics of superconductivity.
\end{abstract}

\section{Introduction}
\label{intro}
 
Ted's first paper on superconductivity, published in 1954 \cite{ted1954}, announced
the   discovery of $Nb_3Sn$, the highest $T_c$ material known since  then and for   two decades thereafter. 
This was a year after he had studied the contribution of electron and hole carriers to the transport properties of Ge \cite{holesge}. In fact, in the period
1953-1955 Ted wrote seven papers dealing with either Seebeck or Hall coefficients of various  materials \cite{holesge,tedholes}, which  indicates that he was very familiar with the concepts of electrons and holes. Unfortunately Ted did not
study the nature of the charge carriers in $Nb_3Sn$ at the time. If he had, he would have learned that the charge carriers are holes 
\cite{holesnb3sn,holesnb3sn2,holesnb3sn3},
and might have wondered: is the high $T_c$ of $Nb_3Sn$ in any way related to this fact?  Yet at that time and for
many decades thereafter, the question whether charge carriers in superconductors are electrons or holes was not a focus of attention of researchers \cite{stewart}.
That suddenly changed in 1987, when Uchida and coworkers \cite{h1}  followed by several others \cite{h2,h3,h4} pointed out and emphasized the fact that the charge
carriers in the newly discovered high $T_c$ cuprates \cite{bm} were holes. 

Since 1954 till the present, a period of 65 years, Ted has  uninterruptedly discovered and investigated a  wide range of superconducting materials, first conventional ones, then
transitioning smoothly into the high $T_c$ realm after 1986. One of the goals of these investigations was to
increase our understanding of the physics of superconductivity by studying how it manifests itself in the different materials,
another goal was  to find superconducting materials with properties ($T_c$ and others) that make them useful for technological applications.
Ted's majestic body of work has contributed greatly to both.

 When  Ted started working on superconductivity, back in the early 50's before  BCS theory existed, I imagine that one of his main goals  was to contribute to the  understanding of ``$the$'' mechanism that leads 
 to this very unusual state of condensed matter. He was not trying to understand ``$a$'' mechanism of superconductivity nor, I  suspect,
 was anybody else at that time. It was only after the advent of BCS theory in 1957, that was quickly generally accepted
 as the correct theory to describe most superconducting materials known at that time, that the question arose in Ted's and other scientists' minds:
 could there be another mechanism besides BCS electron-phonon to explain the superconductivity of
 some materials?
 In other words, could there be more than one mechanism  of  superconductivity in nature?
 
 One important discovery that raised this question was Ted et al's finding in 1961  that $Ru$ exhibits no isotope effect \cite{tedru}.
 Fortunately as Ted relates it \cite{epilogue}, unfortunately as I see it, a few weeks later Ted ran into Phil Anderson at his office
down the hall  at Bell Labs, who convinced him that Ted's important finding was not that $Ru$ was a non-BCS superconductor
 but rather that $Ru$ manages to avoid the BCS isotope prediction because its Coulomb pseudopotential 
 depends on isotopic mass in precisely the right way so as to cancel the BCS-expected dependence of
 $T_c$ on isotopic mass $M$, $T_c\propto M^{-1/2}$, leading to, as Ted puts it \cite{epilogue}, ``a deeper understanding of BCS''.
 Many more such ``deeper understandings'' of BCS would be found in the ensuing years to explain away
 discrepancies between observations and straightforward  BCS, including the negative isotope effects of 
 $U$ \cite{uisotope} and $PdH$ \cite{pdhisotope}. This continued more or less until 1986, the start of the high $T_c$ cuprate era \cite{bm},  when  the scientific  community was finally forced to admit beyond reasonable doubt that
 BCS-electron-phonon could not be the universal explanation for superconductivity in all materials.
 
  In the last 40 years, a large variety of new classes of superconducting materials have been discovered. 
 In a recent Special Issue of Physica C \cite{specialissue}, we compiled  information contributed by leading experimental scientists on 12 classes of 
 `conventional superconductors', i.e. generally agreed to be described by BCS-electron-phonon,  and 20 other classes that are either generally believed to be
 `non-conventional' (i.e. not described by BCS), or where there is doubt whether  they are `conventional'  or not.
 Ted and coworkers contributed a nice introductory article \cite{whattctells} to  this Special Issue giving their Weltanschauung on superconducting materials grounded on his extensive experience.
 
 Ted himself has advocated a non-conventional mechanism for   materials that most consider to be   conventional
 BCS superconductors \cite{whattctells},  such as $PbTe$ doped with a few percent of $Tl$ \cite{pbte} or 
 $Sn_{1-\delta}Te$ doped with $In$ \cite{snte}. According to Ted, $Tl$ and $In$ are  `negative $U$ centers' leading to a 
 much higher $T_c$ in $Tl_xPb_{1-x}Te$ and $Sn_{1-\delta-x}In_xTe$ than would be expected from just the electron-phonon interaction \cite{bustarret}.
 Ted has also proposed that negative $U$ centers contribute to the high $T_c$ of the cuprates \cite{tednegucuprates}.

 So a natural question arises: is there $any$ unifying feature common to all the superconducting materials that
 Ted has investigated, such as $Nb_3Sn$  \cite{ted1954}, Ruthenium  \cite{tedru}, Molybdenum \cite{tedmo}, Iridium \cite{tedir}, $Nb_3Ge$ \cite{tednb3ge}, 
 graphitic compounds \cite{tedgr}, Antimony \cite{tedas}, barium tungsten bronze \cite{tedbtb}, $TaS_2$ \cite{tedtas2}, $PbTe$ \cite{pbte}, 
$Sn_{1-\delta}Te$ \cite{snte},
 hole-doped  cuprates \cite{tedhightc}, electron-doped cuprates \cite{tededoped}, etc?
 I argue  there is: they all have {\it hole carriers}. For most of them it is evident from Hall coefficient measurements or
 valence counting,
 in some cases \cite{tededoped} it may be masked \cite{edoped} by multi-band behavior requiring a more detailed analysis \cite{greene,greven}.
 
 Why are hole carriers conducive to superconductivity? There is to my knowledge not a single paper in the superconductivity literature
 that would argue that the BCS-electron-phonon mechanism favors  holes over electrons.
 Even if it is not yet firmly established experimentally  that superconductivity $requires$ hole carriers, as I have argued for many years \cite{holescfirst,holesc,libro},
 it is clear that in the vast majority of superconducting materials, starting with the periodic table \cite{chapnik},
 the carriers responsible for superconductivity are holes. And,   the highest $T_c$'s in each class of superconductors are realized in materials
 where the carriers are clearly holes, such as $Nb$ \cite{nbholes}, $Nb_3Ge$ \cite{nb3geholes}, 
 $NbSe_2$ \cite{nbse2holes}, $YBa_2Cu_3O_{7-\delta}$ \cite{ybcoholes}, 
 $HgBa_2Ca_2Cu_3O_x$ \cite{hgbaholes}, $MgB_2$ \cite{mgb2holes}, $SmFeAsO_{1-\delta}$ \cite{pnicholes,pnicholes2}.
 Why has there been no effort to explain this remarkable fact?
 
 In the following sections I discuss the fundamental reason for why I  believe  that holes are $indispensable$ for superconductivity in all materials.

 \section{Momentum of the supercurrent in superconductors}
\label{sec:1}
The supercurrent in a superconductor carries mechanical momentum. Consider for simplicity a cylindrical superconductor of radius $R$ and
 height $h$, in an applied magnetic field $H$ parallel to its axis that is uniformly distributed over the cylinder when it is in the normal state.
 In the superconducting state, the magnetic field is expelled from the interior and only penetrates a distance $\lambda_L$, the London
 penetration depth. The electronic mechanical angular momentum associated with the surface current that suppresses the magnetic field in the interior of the
 cylinder is \cite{momentum}
 \beq
\vec{L}_e=-\frac{m_e c}{2e}hR^2 \vec{H}
\eeq
where $m_e$ is the $bare$ electron mass and  $e$ is the electron charge (with its sign). This mechanical momentum has been measured experimentally \cite{gyro}. It  originates in $n_s$ electrons per unit volume flowing within the 
London penetration depth $\lambda_L$ of the surface with velocity given by
\beq
v_s=-\frac{e\lambda_L}{m_ec}H .
\eeq

At a certain temperature $T_c(H)$ that depends on the applied field $H$, the system will undergo a first order
reversible phase transformation to the normal state.  In the normal state no supercurrent flows, therefore the
total electronic angular momentum is zero. Momentum conservation requires that the electronic angular momentum
$\vec{L}_e$, Eq. (1), is transferred to the body as a whole that will start rotating with angular momentum Eq. (1).
If the body is not free to rotate but clamped, the entire earth will acquire the angular momentum Eq. (1).

The current cannot stop simply by onset of resistance, transferring its mechanical momentum to the body by collisions, because  Joule heat would be generated making the process irreversible.
It has been shown experimentally that no Joule heat is generated \cite{keesom}, and we also know from  theory   \cite{gortercasimir,bcs} that the  transition is thermodynamically reversible, hence has to take place without dissipative
processes that would raise the entropy of the universe. Otherwise, basic thermodynamic relations that have been amply verified experimentally such as
Rutger's relation \cite{rutgers}  wouldn't hold.

I argue that only charge carriers with negative effective mass, i.e. holes, can do this \cite{whyholes}.

The   reason is simple to explain. According to Bloch's semiclassical transport theory, negative effective mass means that when an 
external force is applied to the electron, the electron accelerates in direction opposite to the applied force.
This obviously means that there is another force, resulting from the interaction of the electron with the
crystalline array of ions, that pushes in the opposite direction and is larger than the applied external force. 
The fact that there is this force exerted by the lattice on the electrons means that there is momentum
transfer between the electrons and the ionic lattice. And this momentum transfer involves no
scattering or dissipation, it arises from the coherent interaction of the electron wave in a state near the
top of the band with the ions in the crystal.

If instead the carriers are electrons rather than holes, their effective mass is positive and they react to an
external force by accelerating in the direction of the external force. In this case the electron-lattice interaction
does not play a significant role, and there is no net momentum tranfer between electrons and the lattice.

Thus, I argue that if the normal state carriers in a material are only electrons, a supercurrent in that material would not 
be able to stop without violating physical laws, because there would be no mechanism to transfer the momentum of the supercurrent to the lattice without dissipation.
Therefore I conclude that ``electron superconductors'', meaning superconducting materials
that don't have hole carriers, don't exist.

If the above is correct, then BCS theory cannot be $the$ correct theory for conventional superconductors, contrary to what is  generally believed. 
Because within BCS theory, superconductors can exist when the normal state charge carriers are either holes or electrons.
I have argued that BCS lacks essential physical elements to describe superconductivity in nature. In particular, that
it cannot explain the Meissner effect. This is discussed in the next section.

  \section{BCS theory and the Meissner effect}
\label{sec:2}
  In their original paper \cite{bcs}, BCS argued that their theory explains the Meissner effect. This was subsequently questioned because the theory
  violates  gauge invariance,
  but it was then established that this can be fixed \cite{gauge}. However, I argue that BCS  does not explain the Meissner effect for very different reasons.
  
  Let us  review the argument by which BCS (and others) prove that BCS predicts the Meissner effect. They consider the linear response of a system 
{\it in the BCS state} to 
the perturbation created by a magnetic field, as shown in Fig. 1. The perturbing Hamiltonian is the linear term in the
magnetic vector potential $\vec{A}$ that results from
the kinetic energy $(\vec{p}-(e/c)\vec{A})^2/2m$, and has the form
\beq
H_1=  \frac{ie\hbar}{2mc}     \sum_i  (\vec{\nabla}_i\cdot{A}+\vec{A}\cdot\vec{\nabla}) 
\eeq
This perturbation causes the BCS wavefunction $|\Psi_G>$ to become, to first order in $\vec{A}$
\beq
|\Psi>=|\Psi_G>-\sum_n\frac{<\Psi_n|H_1|\Psi_G>}{E_n}|\Psi_n>
\eeq
where  $|\Psi_n>$ are states obtained from the BCS state $|\Psi_G>$ by exciting 2 quasiparticles, and $E_n$ is the excitation energy. The expectation value of the current operator with this wave function gives the electric current $\vec{J}$:
\beq
\vec{J}=<\Psi|\vec{J}_{op}|\Psi>=-\frac{c}{4\pi}K\vec{A}
\eeq
where $K$ is the London Kernel. I have omitted wavevector dependence here for simplicity. In the long wavelength limit
this calculation yields \cite{bcs}
\beq
K=\frac{1}{\lambda_L^2}
\eeq
where $\lambda_L$ is the London penetration depth. Eq. (5) is the (second)  London equation.
In combination with Ampere's law, Eq. (5) predicts that the magnetic field does not penetrate the
superconductor beyond a distance $\lambda_L$ from the surface, where the current $\vec{J}$ circulates,
as shown schematically in Fig. 1 right panel.

          \begin{figure} 
 \resizebox{12cm}{!}{\includegraphics[width=6cm]{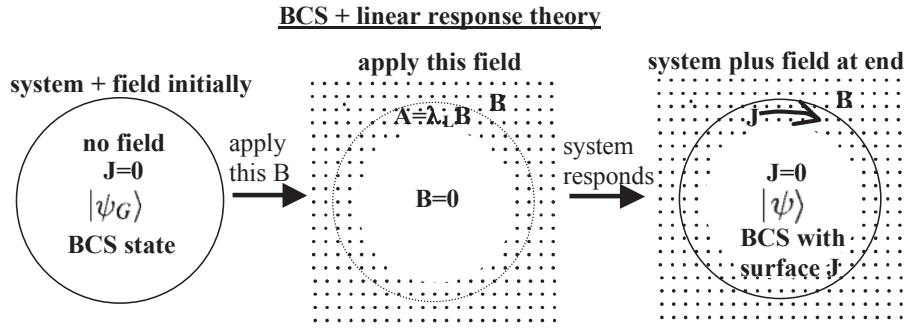}}
 \caption { 
 The BCS view of the Meissner effect. In the BCS explanation of the Meissner effect, the system (cylinder, top view) is in the BCS state (left panel) initially with no magnetic field, and its linear
 response to the magnetic field shown in the middle panel (dots) is computed to first order in the magnetic field. The result
 is the state shown in the right panel, with a surface current $J$ circulating.}
 \label{figure1}
 \end{figure}

            \begin{figure}
 \resizebox{12cm}{!}{\includegraphics[width=6cm]{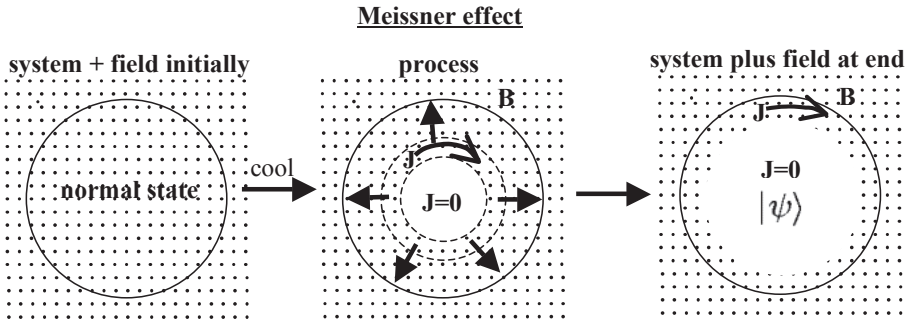}}
 \caption { What the Meissner effect really is: the $process$ by which
 a normal metal becomes superconducting in the presence of a magnetic field throughout its interior initially.
 The simplest route in this process (not the only one) is depicted in the figure.
 The superconducting region (white region) expands gradually  from the center to
 fill the entire volume, expelling the magnetic field in the process. }
 \label{figure1}
 \end{figure} 
 
The calculation just described, which is in essence what BCS, Anderson, Rickayzen and all the other
BCS references do \cite{gauge},  {\it  uses only the BCS  wavefunction in and around the BCS state}, namely the 
ground state wavefunction $|\Psi_G>$ and the wavefunctions $|\Psi_n>$ that result from breaking one Cooper
pair at a time.  The wavefunction of the normal metal never appears in these calculations.

That is $not$ explaining the Meissner effect. The Meissner effect  is what is shown in Fig. 2:  the process by which a system starting in the
normal metallic state expels a magnetic field in the process of becoming a superconductor. It cannot be explained by starting from the assumption that the system is  in the final BCS state and   gets perturbed by $H_1$. Explaining this 
process  requires explaining how the interface between normal and superconducting regions moves (center panel in Fig. 2).
 
Maybe the reader will think: when the system is cooled, the normal state wavefunction somehow turns into the BCS wavefunction, 
and then the perturbing Hamiltonian $H_1$ acts  and the magnetic field gets expelled. However this cannot be so: the BCS state
$|\Psi_G>$ has global phase coherence, and phase coherence cannot exist in the presence of a magnetic field in the interior
of the system. So the system cannot go into the BCS state in the presence of the magnetic field. It has to evolve into the
superconducting state  as it expels the magnetic field.

The  BCS-based calculations do nothing   to explain the Meissner effect
because they do nothing to explain how the system evolves
from the initial state to the final state. Calculations of the sort described in Eqs. (3)-(5) contain no information
 about what is the nature of
the initial state when the Meissner effect starts, the normal metal, so they cannot be a microscopic derivation of the Meissner effect.

There have also been calculations  \cite{dorsey,goldenfeld} of the kinetics of the  transition process using time-dependent Ginzburg-Landau theory  \cite{tdgl}.
That formalism is phenomenological and  
involves a  first order differential equation in time with $real$ coefficients for the time evolution of the order parameter. Hence it describes
$irreversible$ time evolution, and is therefore not relevant to the Meissner effect for type I superconductors,
which is a $reversible$ process  as discussed above. 
There is also theoretical work 
in the literature on the
resistive transition in a magnetic field \cite{resist} describing the onset of resistance in type II superconductors
 through the creation of phase slips at a finite rate. 
However such treatments are also  not relevant to the physical situations discussed above. 

During the process of field expulsion, as well as  its reverse, the process where a superconductor with a magnetic field  excluded turns normal and the field 
penetrates, a Faraday electric field is generated that opposes the process. This electric field drives current in direction opposite to the current that
develops. So it is necessary to explain:
 
1) How can a Meissner current start to flow in direction opposite to the Faraday electric force resisting magnetic flux change (Lenz's law)?
 
2) How is the   angular momentum of the developing supercurrent compensated so that momentum conservation is not violated?
 
3) When a supercurrent stops, what happens to the angular momentum that the supercurrent had?
 
4) How can a supercurrent stop without generation of Joule heat and associated with it an irreversible increase in the entropy of the universe that is known not to occur?
 
 These questions are not addressed in the BCS literature. I have argued in several recent papers that BCS cannot answer these questions, and
 proposed answers to these questions  \cite{ondyn,disapp,revers,whyholes,momentum,entropy}  based on the alternative theory of hole superconductivity
 \cite{holesc}. I have been told in private conversations
 with BCS experts and through referee reports on my papers that alternative theories should not be considered because  these questions  can `in principle' be addressed and answered within
 the BCS framework. I  look
 forward to the day that somebody will actually show that to be the case (or not) by publishing such calculations in the scientific literature, rather than speculating
this could be done if only they didn't have more important things to do.

\section{The key to the Meissner effect}
\label{sec:3}
The key to properly understand the Meissner effect is Faraday's law. Faraday's law tells us that electrical conductors oppose changes in magnetic flux, the more so the
better conductors they are. Superconductors then 
should `superoppose' changes in magnetic flux. How come metals becoming superconducting do the opposite, i.e. expel magnetic flux?

Plasma physicists know the answer. It is embodied in what is known as Alfven's theorem: in a perfectly conducting fluid, magnetic field lines are frozen into the fluid
and hence can only move together with the fluid (if the fluid is imperfectly conducting, magnetic field lines both move with the fluid but can also have some relative motion with respect to the fluid).

So if magnetic field lines move out when a metal becomes superconducting, it is only natural to conclude that a conducting fluid is moving out together with the field lines.
But then how come plasma physicists have not explained the Meissner effect to solid state physicists long time ago?

It's because plasma physicists   don't know about holes.

Plasma physicists know that if a charge-neutral fluid composed of negative and positive charges moves outward, it will carry magnetic field lines with it.
Because it is charge neutral, the flow will not give rise to charge inhomogeneity. However, it will give rise to mass inhomogeneity, since there is
a net outflow of mass.

But we solid state physicists (at least some of us) know that holes are not `real' particles: that when holes are moving in one direction, physical mass and physical mechanical momentum actually move in the opposite direction.
In other words, in solids we can have  flow of negative and positive charge (electrons and holes) in the same direction that does not give rise to either charge nor mass imbalance. In plasmas that is not possible.

How this works in detail is discussed in the references \cite{ondyn,disapp,revers,whyholes,momentum,entropy,alfvenwaves}.

\section{`Conventional' and `unconventional' superconductors}
\label{sec:4}
According to  Wikipedia's page on ``High-temperature superconductivity'', {\it ``the origin of high-temperature superconductivity is still not clear, but it seems that instead of electron-phonon attraction mechanisms, as in conventional superconductivity, one is dealing with genuine electronic mechanisms (e.g. by antiferromagnetic correlations), and instead of conventional, purely s-wave pairing, more exotic pairing symmetries are thought to be involved (d-wave in the case of the cuprates; primarily extended s-wave, but occasionally d-wave, in the case of the iron-based superconductors''}.

There are by now many different classes of `unconventional superconductors' \cite{specialissue}, and it does not seem at all obvious that the same `unconventional'
mechanism can explain all of them. At the same time, it is not clear why, if there are several non-conventional superconductivity mechanisms, they are generally assumed to play
absolutely no role in the so-called `conventional superconductors' such as the elements and simple compounds \cite{webb}.

An alternative scenario is that there is a single mechanism for superconductivity that applies to all materials, and apparent differences between materials
are due to specifics of the materials that are not directly related to their superconductivity. That single mechanism cannot be the BCS-electron-phonon mechanism
because it cannot explain the high $T_c$'s of the cuprates. Therefore, within this alternative scenario the electron-phonon interaction is $not$ the reason 
superconductivity occurs in the conventional superconductors either. The alternative mechanism, if it is able to explain $T_c$'s of 140K in the cuprates, 
should not have much difficulty in  also
  explaining  $T_c$'s of under $10K$ in the elements, as well as  why lattice vibrations can slightly modify, not $cause$, such low temperature superconductivity.

Since 1989, in work principally in collaboration with Frank Marsiglio, we have proposed that hole superconductivity is this alternative mechanism \cite{holesc}.

\section{Theory of hole superconductivity}
\label{sec:5}
  We have proposed that pairing of hole carriers, originating in the Coulomb interaction between electrons in Bloch states, explains superconductivity in
  the cuprates \cite{hole1,hole2,hole3} as well as  in all other 
  superconducting materials \cite{bondch}. Superconductivity is driven by lowering of kinetic \cite{kinetic1,kinetic2}  rather than of potential energy as in BCS. 
  There are distinct experimental signatures of this mechanism such as tunneling asymmetry \cite{tunn} and apparent violation of the optical sum rule \cite{optical}
  that have been verified experimentally well after the theoretical predictions were made.
  
  In the electron-doped cuprates, where the carriers initially were thought to be electrons rather than holes \cite{edoped}, it has by now been convincingly established
  that it is   hole carriers that give rise to the superconductivity \cite{greene,greven,edopedours}.
  
  The theory predicts that highest $T_c$'s result when conduction occurs through holes in negatively charged anions in planes that are negatively charged such as
  the $Cu^{++}-(O^{=})_2$ planes in the cuprates or the $B^-$ planes in $MgB_2$ \cite{matmech}.
  It also explains Mathias' rules \cite{matrules}, a subject that has been of particular  interest to Ted \cite{whattctells}.

\section{Conclusion}
\label{sec:6}
In celebrating this joyous occasion of Ted's coming of age  it is fit to ask: what can theorists  do to help make Ted's search for superconducting materials
  easier in his next 100 years, as well as help others that will tread in Ted's footsteps? Clearly, it is imperative that we develop  clear and  effective
  theoretical criteria to guide the search for new superconductors, something that Bernd Matthias used to complain didn't exist 15 years after BCS
  \cite{matthias}  and still
  doesn't exist today, more than 60 years after BCS. To be effective, those criteria have to be solidly grounded  in the correct theoretical understanding of the
  fundamental physics of superconductors. 
   I have argued \cite{validity} that the fact that BCS-derived criteria have not led to the finding of higher $T_c$ superconductors  
is additional strong evidence  that BCS does not correctly describe the fundamental physics of superconductors.

Based on the theory of hole superconductivity, there are two alternative ways to find high temperature superconductors:

(1) Materials where holes conduct through closely spaced negatively charged anions. Such materials tend to be unstable, for several  reasons: 
(i) packing negative charge costs a lot of Coulomb energy, and (ii) since bands are almost full for hole carriers to exist, a lot of $antibonding$ states are
occupied, and antibonding electrons push to break the solid apart. As Bernd Matthias and Ted learned from experience \cite{persuade,tedinstab},
pushing $T_c$ higher often leads to lattice instabilities \cite{instab}. Overcoming this difficulty  will likely continue to be an art as much  as a science.

(2) Materials with no hole carriers. As explained in Sect. II, a supercurrent in such a material cannot stop
without violating physical laws, hence  once it is created it will continue to flow forever even at room temperature.

Although both ways are difficult, I believe one is infinitely more difficult than the other. 

I would like to close this article by wishing Ted a very happy birthday and
thanking him for his  immense contribution to this field of endeavor.


%
%



\end{document}